\documentclass[fleqn,12pt,twoside]{article}
\usepackage{espcrc1}

\usepackage{graphicx}
\usepackage[figuresright]{rotating}

\newcommand{\AmS}{{\protect\the\textfont2
  A\kern-.1667em\lower.5ex\hbox{M}\kern-.125emS}}

\title{Charged hadron transverse momentum distributions in Au+Au collisions \\
at $\sqrt{s_{_{NN}}} =$ 200 GeV}

\author{Christof Roland for the PHOBOS Collaboration \\*[0.3cm]
\begin{small}
B.B.Back$^1$,
M.D.Baker$^2$,
D.S.Barton$^2$,
R.R.Betts$^6$,
M.Ballintijn$^4$,
A.A.Bickley$^7$,
R.Bindel$^7$,
A.Budzanowski$^3$,
W.Busza$^4$,
A.Carroll$^2$,
M.P.Decowski$^4$,
E.Garc\'{\i}a$^6$,
N.George$^{1,2}$,
K.Gulbrandsen$^4$,
S.Gushue$^2$,
C.Halliwell$^6$,
J.Hamblen$^8$,
G.A.Heintzelman$^2$,
C.Henderson$^4$,
D.J.Hofman$^6$,
R.S.Hollis$^6$,
R.Ho\l y\'{n}ski$^3$,
B.Holzman$^2$,
A.Iordanova$^6$,
E.Johnson$^8$,
J.L.Kane$^4$,
J.Katzy$^{4,6}$,
N.Khan$^8$,
W.Kucewicz$^6$,
P.Kulinich$^4$,
C.M.Kuo$^5$,
W.T.Lin$^5$,
S.Manly$^8$,
D.McLeod$^6$,
J.Micha\l owski$^3$,
A.C.Mignerey$^7$,
R.Nouicer$^6$,
A.Olszewski$^3$,
R.Pak$^2$,
I.C.Park$^8$,
H.Pernegger$^4$,
C.Reed$^4$,
L.P.Remsberg$^2$,
M.Reuter$^6$,
C.Roland$^4$,
G.Roland$^4$,
L.Rosenberg$^4$,
J.Sagerer$^6$,
P.Sarin$^4$,
P.Sawicki$^3$,
W.Skulski$^8$,
S.G.Steadman$^4$,
P.Steinberg$^2$,
G.S.F.Stephans$^4$,
M.Stodulski$^3$,
A.Sukhanov$^2$,
J.-L.Tang$^5$,
R.Teng$^8$,
A.Trzupek$^3$,
C.Vale$^4$,
G.J.van~Nieuwenhuizen$^4$,
R.Verdier$^4$,
B.Wadsworth$^4$,
F.L.H.Wolfs$^8$,
B.Wosiek$^3$,
K.Wo\'{z}niak$^3$,
A.H.Wuosmaa$^1$,
B.Wys\l ouch$^4$\\
$^1$~Argonne National Laboratory, Argonne, IL 60439-4843, USA\\
$^2$~Brookhaven National Laboratory, Upton, NY 11973-5000, USA\\
$^3$~Institute of Nuclear Physics, Krak\'{o}w, Poland\\
$^4$~Massachusetts Institute of Technology, Cambridge, MA 02139-4307, USA\\
$^5$~National Central University, Chung-Li, Taiwan\\
$^6$~University of Illinois at Chicago, Chicago, IL 60607-7059, USA\\
$^7$~University of Maryland, College Park, MD 20742, USA\\
$^8$~University of Rochester, Rochester, NY 14627, USA\\
\end{small}
}

\date{\today}
       
\begin{document}

\maketitle
\begin{abstract}
We present transverse momentum distributions of charged hadrons 
produced in Au+Au collisions at $\sqrt{s_{_{NN}}} =$ 200 GeV. The evolution 
of the spectra for transverse momenta $p_T$ from 0.25 to 5~GeV/c 
is studied as a function of collision centrality over a range from 65 to 344 
participating nucleons. 
We find a significant change of the spectral shape between $p\bar{p}$
and peripheral Au+Au collisions. Comparing peripheral to central Au+Au 
collisions, we find that the yields at the highest $p_T$ exhibit 
approximate scaling with the number of participating nucleons, rather than
scaling with the number of binary collisions.\\
\end{abstract}
\begin{figure}[htb]
\begin{minipage}[t]{77mm}
\includegraphics[width=7.5cm,height=8.5cm]{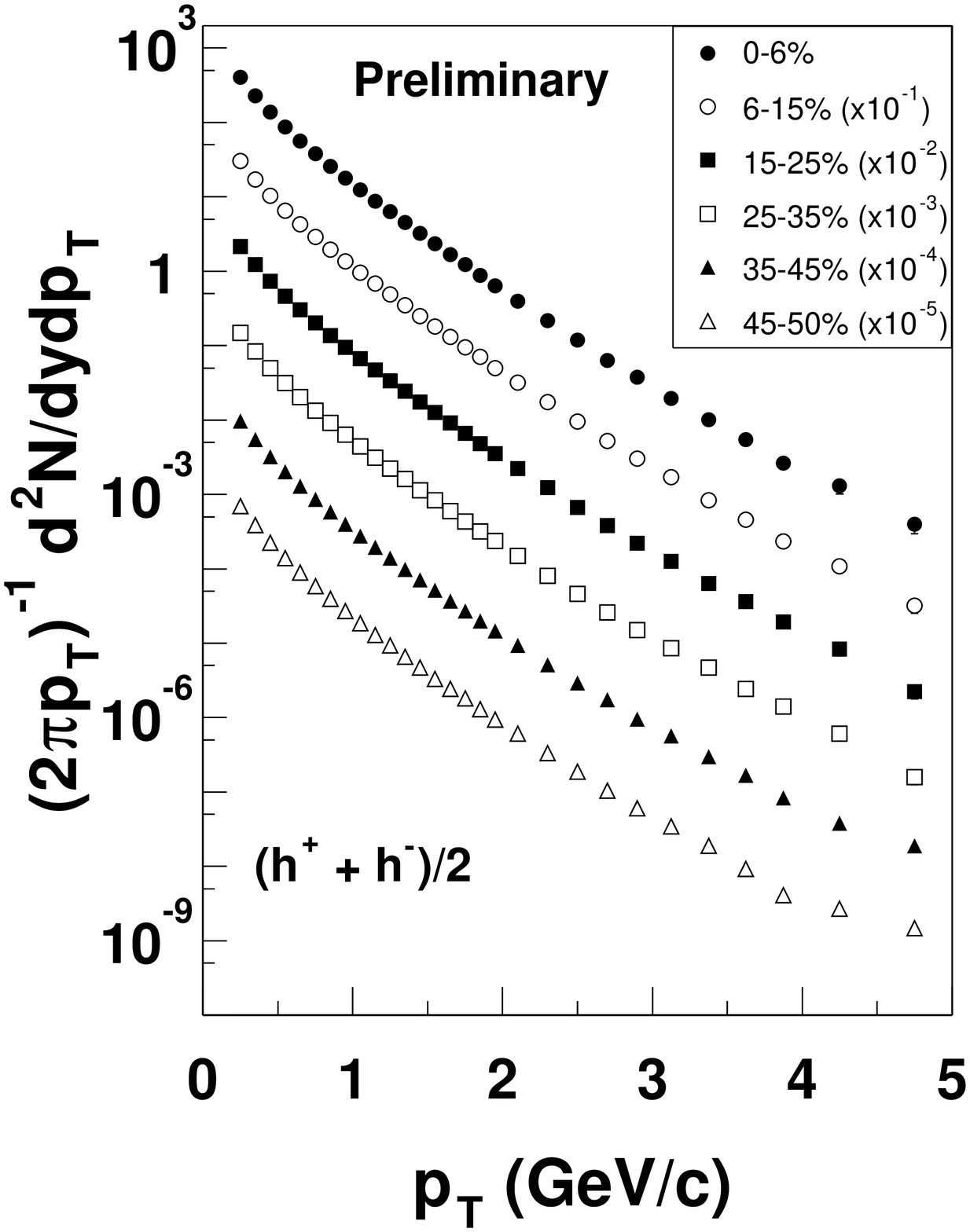}
\vspace{-0.3cm} 
\caption{ \label{ref_SpectraAllCent} 
Invariant yields for charged hadrons as a function of $p_T$ for 6 
centrality bins. For clarity, adjacent bins are scaled by factors of 10.}
\end{minipage}
\hspace{\fill}
\begin{minipage}[t]{77mm}
\includegraphics[width=7.5cm,height=8.5cm]{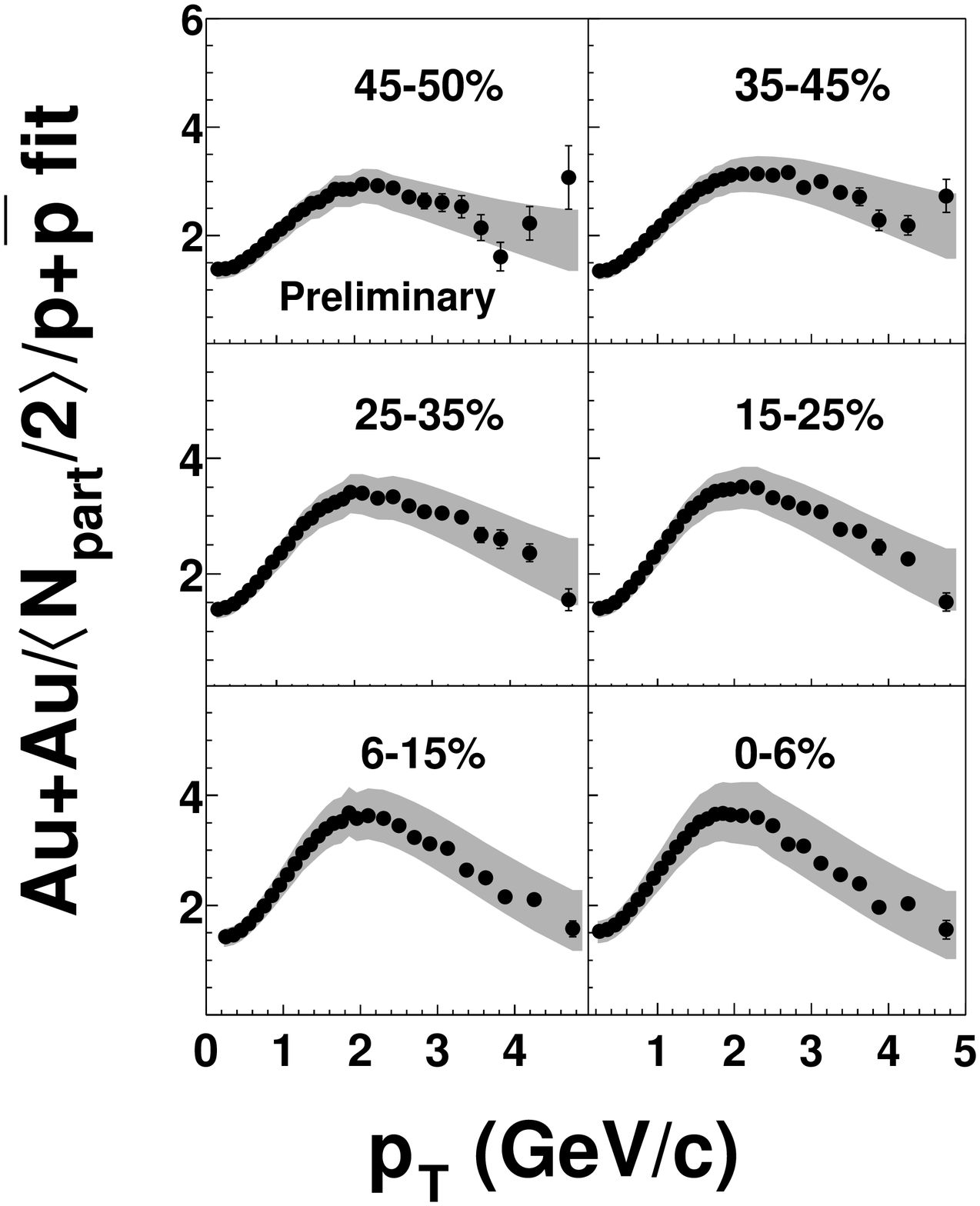}
\vspace{-0.3cm} 
\caption{ \label{ref_SpectraRatioPbarP}
Ratio of the yield of charged hadrons in Au+Au collisions to a fit of proton-antiproton data (see text) 
scaled by $\langle N_{part}/2\rangle$ as a function of $p_T$ for 6 centrality bins. 
}
\end{minipage}
\end{figure}

\begin{figure}[htb]
\begin{minipage}[t]{77mm}

\includegraphics[width=7.5cm,height=8.5cm]{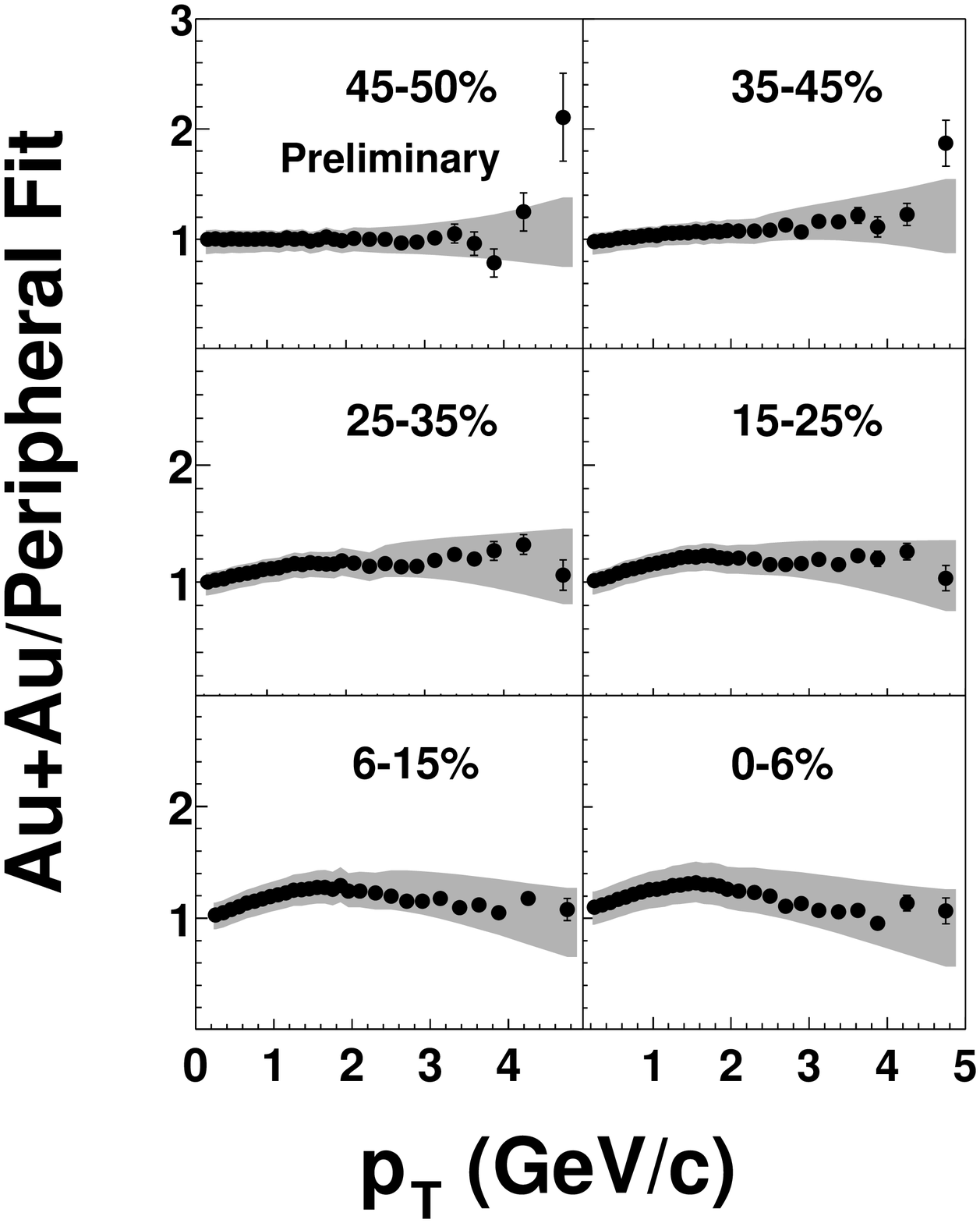}
\vspace{-0.3cm} 
\caption{
\label{ref_RatioPeripheral}
Charged hadron yield in Au+Au in six centrality bins, divided by
a fit to the most peripheral bin and scaled by $\langle N_{part} /2 \rangle$.}
\end{minipage}
\hspace{\fill}
\begin{minipage}[t]{77mm}
\includegraphics[width=7.5cm,height=8.5cm]{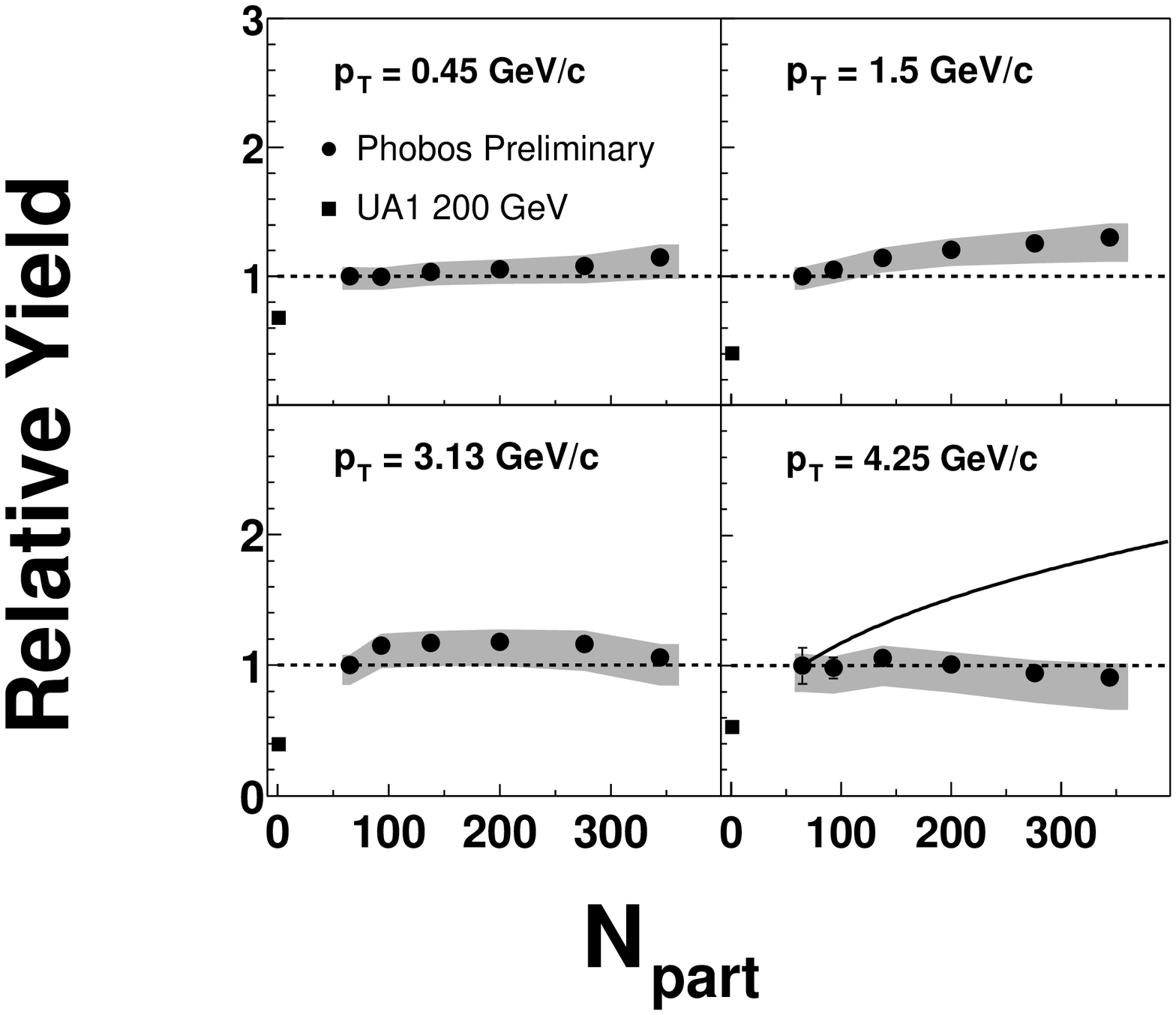}
\vspace{-0.3cm} 
\caption{
\label{ref_YieldCent}
Charged hadron yield in Au+Au scaled by $\langle N_{part}/2\rangle$, normalized to the yield in the most peripheral 
bin, as  a function of $N_{part}$  
at $p_T = $ 0.45, 1.5, 3.13 and 4.25 GeV/c.
}
\end{minipage}
\end{figure}
\begin{table}[htbp]
\begin{center}
\begin{tabular}{|c|c|c|c|c|}
\hline centrality & N$_{events}$ & $ \left<N_{part}\right>$ &  $ \left< N_{coll}\right>$ \\
\hline
\hline
45-50\% & 171,253  & 65   & 107  \\
35-45\% & 173,558  & 93   & 175  \\
25-35\% & 174,121  & 138  & 300  \\
15-25\% & 169,830  & 200  & 500  \\
6-15\%  & 147,074  & 276  & 780  \\
0-6\%   &  73,586  & 344  & 1050 \\
\hline
\end{tabular}
\caption{\label{table1}Details of the centrality bins used. The estimated uncertainty 
in $\langle N_{part} \rangle$ ranges from 
6\% for the most peripheral to 3\% for the most central bin.}
\end{center}  
\end{table} 

%
%

In this paper, the yield of charged hadrons in collisions of gold nuclei at 
an energy of $\sqrt{s_{_{NN}}} = 200$~GeV is presented as a function of 
collision centrality and transverse momentum $p_T$.
The data were taken with the PHOBOS detector 
during the second run of the Relativistic Heavy-Ion Collider (RHIC)
at Brookhaven National Laboratory. 
In the theoretical analysis of particle production in hadronic and nuclear collisions,
a distinction is often made between the relative contributions of ``hard'' parton-parton
scattering processes and ``soft'' processes. 
Hard processes are expected to contribute an increasingly larger fraction of particle production
with increasing collision energy, particularly for produced particles with 
higher $p_T$.

Collisions of heavy nuclei offer ideal conditions to test our understanding 
of this picture, as ``hard'' processes are expected to scale with 
the number of binary nucleon-nucleon collisions $N_{coll}$, whereas ``soft'' 
particle production is expected to exhibit scaling with the number of participating 
nucleons $N_{part}$. In the Glauber picture of nuclear collisions, $N_{coll}$ approximately scales as 
$(N_{part}/2)^{4/3}$. For central collisions of  Au nuclei, one therefore obtains 
an increase in the ratio of $N_{coll}/(N_{part}/2)$ by a factor of six, relative 
to proton-proton collisions. The collision centrality therefore represents a 
control parameter for the relative contributions of hard and soft processes to
particle production.

%
%

The data presented in this paper were collected using the PHOBOS two-arm magnetic spectrometer. 
Details of the setup and track reconstruction procedure can be found elsewhere 
\cite{phobos1,phobos2,highpt,pbarp_200}. 
The invariant cross-sections of charged hadrons were obtained by averaging the yields of 
positive and negative hadrons as a function of transverse momentum for particles with rapidity 
of $0.2 < y_\pi < 1.4$ where the pion mass was assumed in the determination of $y_\pi$.
The cross-sections were corrected for the geometrical acceptance of
the detector, the reconstruction efficiency and the distortion due
to binning and momentum resolution.
We also corrected the spectra for ``ghost'' tracks and particles produced in 
secondary interactions and weak decays.
Details of the corrections applied can be found in \cite{highpt} 

%
%

In Fig.~\ref{ref_SpectraAllCent} the data are shown for 6 centrality bins.  
Table~\ref{table1} gives the event statistics for all centrality bins, along with the average number of participants and binary collisions determined by a Glauber calculation.
Details of the trigger, event selection and centrality determination can be found in \cite{phobos_cent_200}.
The integrated yields, when scaled by $\langle N_{part}/2\rangle$, increase by 15\% over the centrality range, 
in good agreement with the centrality evolution 
of the mid-rapidity particle density presented in \cite{phobos_cent_200}.

To study the evolution of the spectra with centrality we proceed in two steps.
First, in Fig.~\ref{ref_SpectraRatioPbarP} for each centrality bin we show the spectrum
divided by $\langle N_{part}/2\rangle \times f(p_T)$, where $f(p_T)$ is a fit of the invariant 
cross-section in proton-antiproton collisions at the same energy \cite{ua1_pbarp}. 
This comparison shows that already in Au+Au collisions with an impact parameter 
of $b \approx 10$~fm, the spectral shape is modified relative to that in $p+\bar{p}$ collisions. 
It is worth noting that the ratio $\frac{\langle N_{coll}\rangle}{\langle N_{part} / 2\rangle}$ 
increases by a factor of almost three from $p+\bar{p}$ to the most peripheral Au+Au collisions.
Going from peripheral to central events we observe only a moderate change in spectral shape.

The detailed evolution of the spectra with collision centrality
is shown in Fig.~\ref{ref_RatioPeripheral}, where the spectra for the six centrality 
bins have been divided by a fit to the most peripheral bin scaled by $\langle N_{part} / 2\rangle$.
It is remarkable that the change in spectral shape over 
this range of centralities is small compared to that between peripheral events 
and $p+\bar{p}$ collisions. 

In particular at high $p_T > 3$~GeV/c, to a good approximation the yields scale 
with $\langle N_{part} / 2\rangle$ as a function of centrality.
The scaling behaviour in various regions of $p_T$ is further illustrated in Fig.~\ref{ref_YieldCent}. 
Here we show the yield of charged hadrons scaled with $\langle N_{part} / 2\rangle$ relative 
to the yield in the most peripheral bin at $\langle N_{part}\rangle = 65$ 
in Au+Au and proton-antiproton collisions at $p_T  = $ 0.45, 1.5, 3.13 and 4.25 GeV/c. 
The ratio was determined by interpolating the data shown in 
Fig.~\ref{ref_SpectraAllCent} using power law fits.
The centrality range shown in Fig.~\ref{ref_YieldCent} covers a change 
in $\frac{\langle N_{coll}\rangle}{\langle N_{part} / 2\rangle}$ by 
about a factor of two as indicated by the line drawn in the lower right hand panel. 

%
%

The observed trends can be contrasted with the expectation that 
particle production should be characterized by a change from  $\langle N_{part}\rangle$-scaling at low $p_T$ to $\langle N_{coll}\rangle$-scaling at high $p_T$.
No corresponding increase in particle 
production per participant at $p_T = 3$~GeV/c and above is observed. Rather,
the yields in this region scale approximately with the number of participating nucleons.
It has recently been argued that the observed scaling could be naturally 
explained in a model assuming the dominance of surface emission of 
high $p_{T}$ hadrons \cite{mueller}.
However, the approximate participant scaling has also been predicted in
the context of initial state saturation models \cite{dima}.
Upcoming studies of d+Au collisions at RHIC should also provide
insight into the modification of particle spectra in a nuclear environment.\\

This work was partially supported by US DoE grants DE-AC02-98CH10886,
DE-FG02-93ER-404802, DE-FC02-94ER40818, DE-FG02-94ER40865, DE-FG02-99ER41099, W-31-109-ENG-38.
NSF grants 9603486, 9722606 and 0072204. The Polish groups were partially supported by KBN grant 2-P03B-10323. The NCU group was partially supported by NSC of Taiwan under 
contract NSC 89-2112-M-008-024.

\end{document}